\documentclass[sigconf, nonacm]{acmart}
\AtBeginDocument{%
  }

\acmConference[SCORED '26]{ACM Conference on Software Supply Chain Offensive Research and Ecosystem Defenses}{October 6, 2026}{Prague, Czechia}

\usepackage{tabularx}
\usepackage{multirow}
\usepackage{array}
\usepackage{bm}

\begin{document}

\title{AgOSS: A Dataset and Multi-Layer Characterization of Open-Source Agricultural Software}
\subtitle{Agricultural Open-Source Software Security Is a Maturity Problem, Not a Domain Problem}

\author{Vatsal Dudhaiya}
\email{vdudhaiy@purdue.edu}
\affiliation{%
  \institution{Purdue University}
  \city{West Lafayette}
  \state{Indiana}
  \country{USA}
}

\author{Mikhail Golovenchits}
\email{mgoloven@purdue.edu}
\affiliation{%
  \institution{Purdue University}
  \city{West Lafayette}
  \state{Indiana}
  \country{USA}
}

\author{Aryan Banerjee}
\email{banerj61@purdue.edu}
\affiliation{%
  \institution{Purdue University}
  \city{West Lafayette}
  \state{Indiana}
  \country{USA}
}

\author{James C. Davis}
\email{davisjam@purdue.edu}
\affiliation{%
  \institution{Purdue University}
  \city{West Lafayette}
  \state{Indiana}
  \country{USA}
}


\begin{abstract}
Much of agriculture depends on open-source software spanning farm management platforms, cloud services, edge gateways, embedded systems, and field-deployed sensors, forming a domain-specific software supply chain that has drawn little empirical security attention. It is unknown whether this ecosystem's supply chain security posture differs from that of comparable non-agricultural software, and if it does, whether the difference reflects the agricultural domain or the size and maturity of the projects within it.

As a step towards securing agricultural open-source software, we present \textit{AgOSS}, a dataset of 66 repositories across six architectural categories. We assess supply chain security within the dataset via OpenSSF Scorecard, governance metrics, SBOM-based dependency analysis, and KEV matching, and compare against matched non-agricultural controls. We report two findings. First, governance is largely independent of inherited dependency risk. Scorecard tracks community activity, but we detect no association with vulnerability count, density, or known-exploited count, and in regression the exposure signal loads on architectural category rather than domain. Second, agricultural projects score far lower on raw Scorecard, but size and maturity confound the gap: the residual loses significance under matching and regression. Securing this ecosystem means investing in contributor capacity and dependency hygiene, not agriculture-specific controls.
\end{abstract}




\maketitle


\section{Introduction}
Domain-specific open-source software (OSS) ecosystems are increasingly critical to modern infrastructure~\cite{jiang2022,yu2026security}, yet they remain notoriously difficult to define, isolate, and measure~\cite{githubfindings,BaltesSebastian2022}. Agriculture relies on a complex software stack stretching from field-deployed sensor firmware to cloud-hosted farm management platforms~\cite{barreto2018survey,gupta2020security}. While OSS is prevalent throughout this domain~\cite{jsan12020028}, the broader agricultural OSS ecosystem has not been systematically characterized as a distinct software supply chain.

Existing studies primarily focus on domain-specific farm management or modelling applications~\cite{jsan12020028,holzworth2015}, providing limited coverage of the infrastructure layers beneath them. Consequently, the agricultural OSS ecosystem remains poorly defined. Little is known about its supply chain security posture: the adoption of modern security practices, repository governance characteristics, and inherited dependency risk.
We address this gap by asking:
  (1) whether agricultural repositories differ from non-agricultural ones on security and maintenance characteristics;
  (2) how that posture distributes across the stack;
  and
  (3) whether any difference reflects the domain or project maturity.
To answer these questions, we conduct a Mining Software Repositories (MSR) study of the agricultural OSS ecosystem by combining repository governance metrics, OpenSSF Scorecard~\cite{zahan2023scorecard} analyses, GitHub repository metadata, Software Bills of Materials (SBOMs), dependency vulnerability analysis using the Open Source Vulnerabilities (OSV) database~\cite{osv_database}, and Known Exploited Vulnerability (KEV) data from the U.S. Cybersecurity and Infrastructure Security Agency (CISA)~\cite{cisa_kev}.

This paper provides the following contributions:
\begin{itemize}
    \item We present AgOSS, a dataset of 66 open-source repositories mapped onto a four-layer stack model running from cloud-based farm management systems down to field-deployed sensors and embedded infrastructure.
    \item We characterize supply chain security across these repositories and report two findings. First, OpenSSF Scorecard tracks community maturity but we detect no association with vulnerability counts or densities we measure, which vary instead by architectural category. Second, the lower Scorecard results of agricultural projects are confounded by size and maturity: the raw gap does not survive Mahalanobis matching or size-adjusted regression, and contributor count is the variable that most consistently tracks governance.
\end{itemize}

\textbf{Significance:} These results redirect where defensive efforts belong. Hardening governance checks will not reduce inherited dependency risk, since the two move independently. And the apparent agricultural security deficit is a resourcing problem shared with comparably sized projects, so interventions should target contributor capacity rather than the domain itself. That caution generalizes: any domain-level security comparison run without maturity controls risks the same misattribution.


\section{Background and Related Work}
\label{sec:background}

\subsection{Open-Source Software in Agriculture}

Modern agriculture has evolved into a data-driven, interconnected paradigm known as ``smart farming'' or ``precision agriculture.'' The precision farming market was valued at \$15.1 billion in 2025 and is projected to exceed \$38.8 billion by 2033~\cite{grandview2026}. In the United States, automated guidance is applied to over half the acreage planted to major crops (\textit{e.g.}, corn, cotton, rice, etc.) while the more data-intensive practices (\textit{e.g.}, yield maps, soil maps, variable-rate application) reach only 5 to 25 percent of planted acreage~\cite{mcfadden2023precision}. This infrastructure relies on a multi-layered software stack spanning cloud services, edge devices, and field equipment~\cite{barreto2018survey,gupta2020security}.

The agricultural sector is increasingly adopting OSS to reduce licensing costs, mitigate vendor lock-in, improve transparency, and enable low-cost deployment of precision technologies~\cite{jsan12020028}. 
For example, farmOS, a farm record-keeping platform developed by farmers, researchers, and organizations, is catalogued by the Food and Agriculture Organization (FAO) of the United Nations as a tool for producers at all scales~\cite{fao_farmos}. Many beneficiaries of such tools sit outside industrialized agriculture: 72\% of the world's 570 million farms operate on less than one hectare~\cite{lowder2016}, and OSS is one of the few affordable routes to digital tooling for them~\cite{chandra2021digital}. 
Agriculture is not the first sector to undergo this transition. Automotive infotainment consolidated onto Linux-derived stacks shipping in production vehicles~\cite{agl}, and researchers later found supply chain vulnerabilities below the application layer in such firmware~\cite{yu2026security}. 
Likewise, open machine learning model hubs became shared infrastructure before their supply chain risks were characterized~\cite{jiang2022}; in July 2026, a malicious dataset uploaded to one such hub gave attackers code execution on its processing infrastructure~\cite{hf2026incident}.

The agricultural OSS ecosystem spans farm management platforms, telemetry systems, embedded RTOS, edge gateways, cloud dashboards, and data processing libraries. Unlike general-purpose software ecosystems that benefit from large maintainer communities and mature software supply chain security practices~\cite{zahan2023scorecard}, these projects are typically maintained by fragmented networks of researchers, open-source contributors, and small development teams~\cite{jsan12020028}. That fragmentation does not diminish the software's importance. Three groups depend on it: small and mid-sized producers and the public institutions that serve them, for whom free tools are often an affordable route to digital record-keeping~\cite{chandra2021digital,fao_farmos}; researchers and extension programs that use these platforms to collect field data~\cite{fao_farmos}; and commercial operators, who may never deploy open-source software directly but still ship products built on open-source foundations~\cite{ossra2026}, including real-time operating systems, gateway servers, and drone autopilots. Fragmented maintenance and broad dependence introduce largely unmeasured supply chain risks. Failures anywhere in the stack can result in hardware downtime, wasted inputs, and reduced crop yields~\cite{barreto2018survey,gupta2020security},  making the security posture and dependency health of these repositories critical.

\subsection{Related Work By Topic}
\label{sec:rel-topics}
To the best of our knowledge, only one prior study maps OSS within agriculture, and it does so at the level of application availability and feature categorization. Specifically, dos Santos \textit{et al.}~\cite{jsan12020028} conducted a rapid review which cataloged and evaluated 21 open-source precision agriculture packages across application paradigms (\textit{e.g.}, automation, monitoring, and hardware/IoT integration), positioning open-source alternatives as highly customizable, low-cost alternatives to proprietary farming systems. While the authors note benefits such as algorithmic transparency, their evaluation also flags reliance on small, fragmented maintainer communities and weak repository management across the tools they surveyed~\cite{jsan12020028}. Adjacent work reviews the agricultural modelling-software landscape more broadly~\cite{holzworth2015}, but does not treat it as an open-source ecosystem or examine its supply chain. Those governance weaknesses give reason to expect weaker security practices than in mainstream OSS. Whether such a gap reflects the agricultural domain itself or the maturity conditions that happen to accompany it is the question this study addresses. With a single prior mapping to build on, no established corpus, taxonomy, or baseline exists for agricultural OSS, so this study is necessarily exploratory; we construct all three.
 
\subsection{Related Work By Methodology}
\label{sec:rel-methods}
A substantial body of research in empirical software engineering has established methodologies for MSR to study software evolution, dependency structures, and ecosystem-level properties. To ensure empirical rigor, contemporary MSR research heavily scrutinizes pipeline design and domain specialization. Tutko \textit{et al.}~\cite{tutko2022} evaluated workflows, dataset selection, and reproducibility across 286 empirical papers, revealing a widespread lack of standardized reporting metrics and pipeline documentation. These identified gaps directly inform the reproducible pipeline architecture and strict mapping protocols used to build the AgOSS dataset. Complementing this, Soliman \textit{et al.}~\cite{soliman2025} demonstrated through a systematic mapping study of architectural MSR research that mining methodologies must be specialized to specific analytical goals rather than broadly applied. By that reasoning, agricultural computing should also be viewed as a unique software ecosystem requiring specialized analysis of its distinct operational stack.
 
Beyond general frameworks, empirical studies of software supply chains in adjacent domains provide concrete precedents for analyzing multi-layer ecosystems. Jiang \textit{et al.}~\cite{jiang2022} modeled the pre-trained machine learning model supply chain, demonstrating how to map heterogeneous repositories and operationalize supply chain risk through measurable repository attributes. Similarly, Yu \textit{et al.}~\cite{yu2026security} executed a security risk assessment of the Android Automotive OS (AAOS) ecosystem, utilizing firmware reverse engineering and Software Bill of Materials (SBOM) generation to uncover vulnerabilities across deep software layers. Both studies establish that capturing domain-specific risk requires evaluating an entire operational stack—from low-level firmware to user-facing applications. We adopt the same stack-wide scope, translating its risk metrics into the agricultural software domain.
 
Finally, Zimmermann \textit{et al.}~\cite{zimmermann} modeled the npm JavaScript ecosystem as a ``small world'' network of densely connected dependencies and showed both that a small set of maintainers is implicitly trusted across the ecosystem and that vulnerable code persists downstream long after patches ship (40\% of packages). While their focus was on web ecosystems, this network-driven approach to tracking dependency risk and maintainer concentration serves as the primary structural inspiration for the multi-layer dependency analysis implemented in this paper.
 
Closest to our work, Zahan \textit{et al.} measured OpenSSF Scorecard adoption across npm and PyPI~\cite{zahan2023scorecard} and tested whether Scorecard scores predict vulnerability counts, finding weak, counterintuitively positive associations~\cite{zahan2023practices}; follow-on work applied propensity score matching to isolate practice-level effects~\cite{zahan2025prioritizing}. Parallel work has traced how vulnerable dependencies propagate through package networks~\cite{decan2018impact} and how they affect downstream projects~\cite{prana2021outofsight}. These studies are confined to web package ecosystems; whether their findings hold in cyber-physical stacks spanning firmware and RTOS layers is untested — a gap our paper addresses.
 
Table~\ref{tab:gaps} summarizes the gaps identified in \S~\ref{sec:rel-topics} and \S~\ref{sec:rel-methods}.
 
\begin{table}[t]
\caption{Gap Analysis}
\label{tab:gaps}
\footnotesize
\setlength{\tabcolsep}{4pt}
\renewcommand{\arraystretch}{0.95}
\begin{tabularx}{\columnwidth}{@{}>{\raggedright\arraybackslash}p{2.2cm}X@{}}
\toprule
\textbf{Prior work} & \textbf{What our work contributes} \\
\midrule
MSR method \& rigor~\cite{tutko2022,soliman2025} &
Applies these sampling and reporting conventions to agriculture; releases the pipeline and corpus. \\
Adjacent-domain supply chains~\cite{jiang2022,yu2026security} &
Extends the same stack-wide treatment to agriculture, from RTOS and firmware to farm management platforms. \\
Dependency \& maintainer risk~\cite{zimmermann,zahan2023practices,decan2018impact} &
Tests whether these findings hold outside web package ecosystems, in a stack reaching embedded layers. \\
Practice-effect matching~\cite{zahan2025prioritizing} &
Matches against non-agricultural controls to separate domain from project maturity. \\
Ag OSS tool review~\cite{jsan12020028} &
Measures what the catalog did not: dependency exposure, governance, and known-exploited vulnerabilities across six categories. \\
\bottomrule
\end{tabularx}
\end{table}


\section{Research Questions}
In light of this knowledge gap, we ask:
\begin{itemize}
    \item \textbf{RQ1:} Do ag-specific repositories within the AgOSS dataset exhibit weaker security and maintenance characteristics than non-ag-specific repositories?
    \item \textbf{RQ2:} How is the security posture of agricultural OSS distributed across architectural layers?
    \item \textbf{RQ3}: To what extent do these differences reflect the agricultural domain itself, versus project maturity?
\end{itemize}

Answering them shows whether security efforts should target agricultural software specifically or the resourcing and maintenance gaps it shares with comparable projects, and whether domain-specific security studies can draw valid conclusions without controlling for project maturity.


\section{Methodology}
This study adopts an MSR methodology to empirically analyze the OSS supply chain used in agricultural systems. 

\subsection{Dataset Construction}
\begin{figure*}
    \centering
    \includegraphics[width=0.8\textwidth]{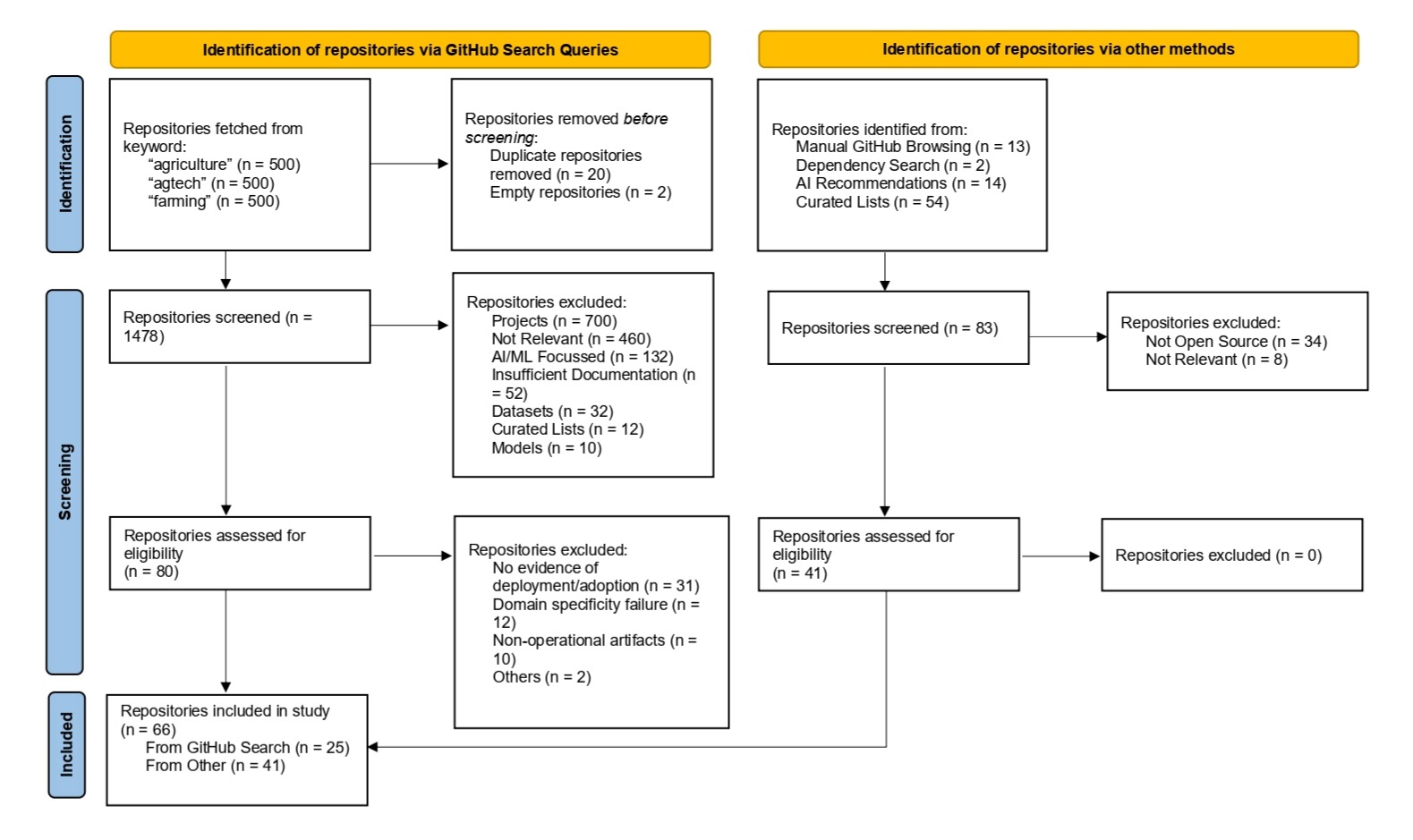}
    \caption{PRISMA flow for AgOSS dataset construction: 1{,}500 keyword candidates, extended by snowball and semantic discovery, screened to 66 repositories in six categories---including the non-ag-specific infrastructure the agricultural stack runs on.}
    \Description{PRISMA Flow Diagram for dataset construction showcasing the identification, screening, and finally included repositories (each tagged with their relevant source and count.}
    \label{fig:prisma_flow}
\end{figure*}

This study employs an exploratory empirical software engineering design to construct and analyze a dataset of agricultural open source repositories. Figure~\ref{fig:prisma_flow} presents the PRISMA flow diagram~\cite{prisma} for dataset construction: identification, screening, eligibility assessment, and inclusion of repositories.

\subsubsection{Search Protocol}
The GitHub Search API was used as follows:
\begin{itemize}
    \item \textbf{Queries:} Primary keyword searches included ``agriculture'', ``agtech'', and ``farming''.
    \item \textbf{Sorting:} Results were sorted using GitHub’s default relevance-based “Best Match” ordering to prioritize repositories most strongly associated with the query terms.
    \item \textbf{Candidate Pool:} We retrieved the top 500 repositories per keyword (total $n=1500$) ranked by GitHub’s default relevance ordering. To ensure quality and reduce duplication, we excluded forked or archived repositories.
\end{itemize}

GitHub keyword-based retrieval has known limitations~\cite{githubfindings}, which is why we also conducted iterative snowball sampling~\cite{BaltesSebastian2022}, including extraction of repositories from curated agricultural OSS lists~\cite{awesome_agriculture,awesome_open_ag,opensourceagriculture} and adjacent geospatial directories~\cite{awesome_geospatial_companies}.

To improve coverage and reduce keyword mismatch, we supplemented manual snowball sampling with LLM-assisted candidate discovery using Claude Sonnet 5. The model was given (1) the domain context of the study, and (2) the candidate repositories discovered up to the termination of the manual keyword search, then prompted to suggest missing, structurally relevant OSS projects. Every candidate was audited against our inclusion and exclusion criteria (\S~\ref{sec:incl-excl-criteria}) prior to integration. The LLM did not dictate inclusion; it only generated candidates. Reflecting this filtering, only 8 repositories (12\% of the dataset) originated from LLM suggestions.

\subsubsection{Inclusion and Exclusion Criteria}
\label{sec:incl-excl-criteria}
Repositories were evaluated using explicitly defined inclusion and exclusion criteria found in Table~\ref{tab:inclusion_exclusion} to reduce subjective selection bias. 
 
\begin{table*}[t]
\centering
\caption{Inclusion and exclusion criteria for repository selection}
\label{tab:inclusion_exclusion}
\footnotesize
\begin{tabular}{@{}p{0.47\textwidth} p{0.47\textwidth}@{}}
\toprule
\textbf{Inclusion Criteria} & \textbf{Exclusion Criteria} \\
\midrule
\textbf{Domain Specificity:} Repository can be Ag-Specific, defined as software whose primary documented use case is within the agricultural domain.
& \textbf{Non-Operational Artifacts:} No personal projects, course assignments, and proof-of-concept demonstrations lacking downstream adoption. \\
\addlinespace
\textbf{Operational Criticality:} When domain specificity is absent, we accept non-agriculture-specific substrates (\textit{e.g.}, RTOS, embedded Linux distributions) that provide the essential execution environment for field-deployed agricultural logic.
& \textbf{General Infrastructure:} General-purpose infrastructure software, such as desktop operating systems or general-purpose compilers, is excluded to keep the stack focused on agricultural operations. \\
\addlinespace
\textbf{Deployment Evidence:} Project must show evidence of practical deployment, experimental implementation in research-derived systems, or backing by an organization.
& \textbf{Research Artifacts:} Exclude repositories whose primary deliverable is a model, dataset, or benchmark rather than deployable operational software. \\
\addlinespace
\textbf{License Availability:} Repository must make source code publicly available. We adopt an inclusive definition of ``open source'' spanning both OSI-approved licenses and source-available licenses.
& \\
\addlinespace
\textbf{Activity Markers:} Indicators such as commit activity, releases, packages, stars, forks, and associated project websites were used as supporting signals of engagement.
& \\
\bottomrule
\end{tabular}
\end{table*}

\subsubsection{Screening}
Screening followed a funnel:
\begin{itemize}
    \item \textbf{Rapid Visual Triage:} The first author screened the results against repository title, description, and metadata. To bound false negatives at this low-information stage, triage was deliberately
    inclusion-biased: a candidate was discarded only when clearly out of scope (\textit{e.g.}, non-software, tutorials, personal projects); any ambiguous case was promoted to the manual content audit rather than dropped.
    \item \textbf{Manual Content Audit:} High-relevance candidates were subjected to an in-depth manual analysis of their documentation to confirm functional placement.
\end{itemize}

66 repositories satisfied all criteria. Identification ceased once every candidate returned by the search strategies had been evaluated. The complete list, with category coding validation, is in the dataset repository (see \textit{Data and Artifact Availability}).

\subsubsection{Labeling}
\label{sec:labels}
Each selected repository was labeled along two axes: Stack Category and Ag-Specificity. 

Our stack categories are a specialization of the general layered IoT architecture mentioned in \S~\ref{sec:background} for the agricultural domain. Each repository was assigned to exactly one of the following:
\begin{enumerate}
    \item Embedded OS substrate ($n = 5$)
    \item Field-Deployed Sensor ($n = 14$) 
    \item Edge and Gateway software ($n = 13$)
    \item Cloud-hosted backends and dashboards ($n = 19$)
    \item Domain-specific agricultural platform ($n = 9$)
    \item Data Processing Libraries/Tools ($n = 6$)
\end{enumerate}

Each repository also received one of two ag-specificity labels:
\begin{enumerate}
    \item \textbf{Ag-Specific:} Repository designed for use in agriculture.
    \item \textbf{Non-Ag-Specific:} General-purpose repository not designed for agriculture, but occupies a necessary layer in the agricultural stack (\textit{e.g.}, RTOS, embedded Linux distributions).
\end{enumerate}

The repository classifications were assigned by the first author and reviewed by a second reviewer for consistency with the classification criteria. Two discrepancies were identified during the review process (96.97\% agreement, 64/66) and were resolved through discussion to produce the final labels.

\begin{figure}
    \centering
    \includegraphics[width=\columnwidth]{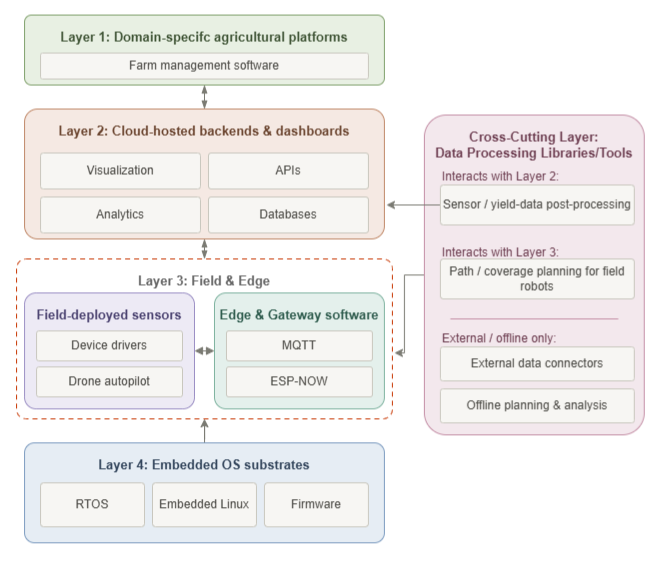}
    \caption{AgOSS stack constructed by the authors, specializing the layered smart-farming architectures described in prior surveys~\cite{barreto2018survey,gupta2020security} into four vertical layers plus one cross-cutting category. Category assignments are our own (\S~\ref{sec:labels}).}
    \Description{Data Flow Diagram for the AgOSS stack with 6 categories: 5 organized into 4 vertical layers, plus a cross-cutting data processing category.}
    \label{fig:stack}
\end{figure}

The ecosystem model (Figure~\ref{fig:stack}) we constructed to organize the corpus is as follows:
\begin{enumerate}
    \item \textbf{Layer 4 (Embedded OS Substrates):} Layer 4 includes Real-Time Operating Systems (RTOS), Embedded Linux, and low-level firmware, and provides the hardware resource management required to run applications on Layer 3 devices. 
    \item \textbf{Layer 3 (Field \& Edge):} Layer 3 is sub-divided into Field-Deployed Sensors (which run device drivers and drone autopilots) and Edge \& Gateway Software (which manage local networks using protocols like MQTT and ESP-NOW).
    \item \textbf{Layer 2 (Cloud-Hosted Backends \& Dashboards):} Layer 2 stores the incoming telemetry in databases, performs large-scale analytics, renders visualizations, and exposes APIs that are consumed by Layer 1.
    \item \textbf{Layer 1 (Domain-Specific Agricultural Platforms):}  This layer houses the user-facing Farm Management Software, which relies on the continuous data flow and structural integrity of the underlying layers to function effectively.
\end{enumerate}

Complementing this four-layer telemetry pipeline is a cross-cutting category,
\textbf{Data Processing Libraries/Tools}, which we deliberately do
not number as a fifth vertical tier. As shown in Figure~\ref{fig:stack}, repositories in this category do not occupy a single fixed position in the sensor-to-platform data flow; instead, they interface with the stack at multiple, heterogeneous points.

\subsubsection{Matched Comparisons}
Observational comparisons of two repository subgroups confound the attribute of interest with correlated differences in programming language, size, age, and activity, which independently predict security outcomes~\cite{zahan2025prioritizing}. Matching addresses this by pairing each repository with controls of similar size and maturity, so that the remaining difference is associated with the characteristic under study rather than with scale~\cite{austin2011introduction}. Following this approach, we match each ag-specific repository to non-agricultural controls of comparable size, age, and activity, and estimate the adjusted domain difference within matched pairs.

\label{sec:matched-methodology}
\paragraph{\textbf{Matching Criteria:}} 
The ag-specific ($n=55$) repositories were matched against control-pool candidates using coarsened exact matching on primary programming language~\cite{iacus2012causal} followed by Mahalanobis distance caliper matching~\cite{mahalanobis1936generalised, rubin1980bias} on six covariates: log-transformed star count, log-transformed fork count, repository age, log-transformed contributor count, log-transformed commit activity over the trailing 52 weeks, and log-transformed codebase size (total non-notebook language bytes was used as a proxy for repository size rather than computing the total lines of code). The caliper threshold was set from the chi-squared distribution with degrees of freedom equal to the number of covariates. The primary matched-pair estimate paired each dataset repository with its $k = 3$ nearest eligible controls by Mahalanobis distance. As a robustness check, $k$ controls were instead drawn at random from the same eligible pool across 1{,}000 seeds, to test whether the primary result depended on always selecting the closest available control rather than on the matching design itself. Covariate balance was assessed via standardized mean differences (SMDs)~\cite{austin2011introduction,stuart2010matching} before and after matching, and results were checked for sensitivity to caliper width (25th/50th/75th percentile) and to k (1, 3, 5).

\paragraph{\textbf{Control Pool Construction:}}
We constructed a pool of non-agricultural comparison repositories through an iterative, three-wave GitHub Search API process. Each wave was guided by post-matching covariate-space coverage diagnostics, including covariate balance, unmatched treated repositories, matching distance distributions, and shortages of eligible controls within specific regions of the covariate space. Candidates were screened using the same inclusion and exclusion criteria described in Table~\ref{tab:inclusion_exclusion}, with the exception of the domain specificity requirement. Across three waves (16 general non-ag terms; 9 SaaS/CRM/ERP terms added when SaaS-niche repos found no controls; 6 Rails / community terms added for the Ruby-language shortfall), we reviewed 1,129 candidates and accepted 649 (463, 106, 80 from waves 1, 2, 3 respectively).

\subsection{Data Pipeline}
\paragraph{Data Collection}
\label{data-pipeline}

\begin{figure*}
    \centering
    \includegraphics[width=0.8\textwidth]{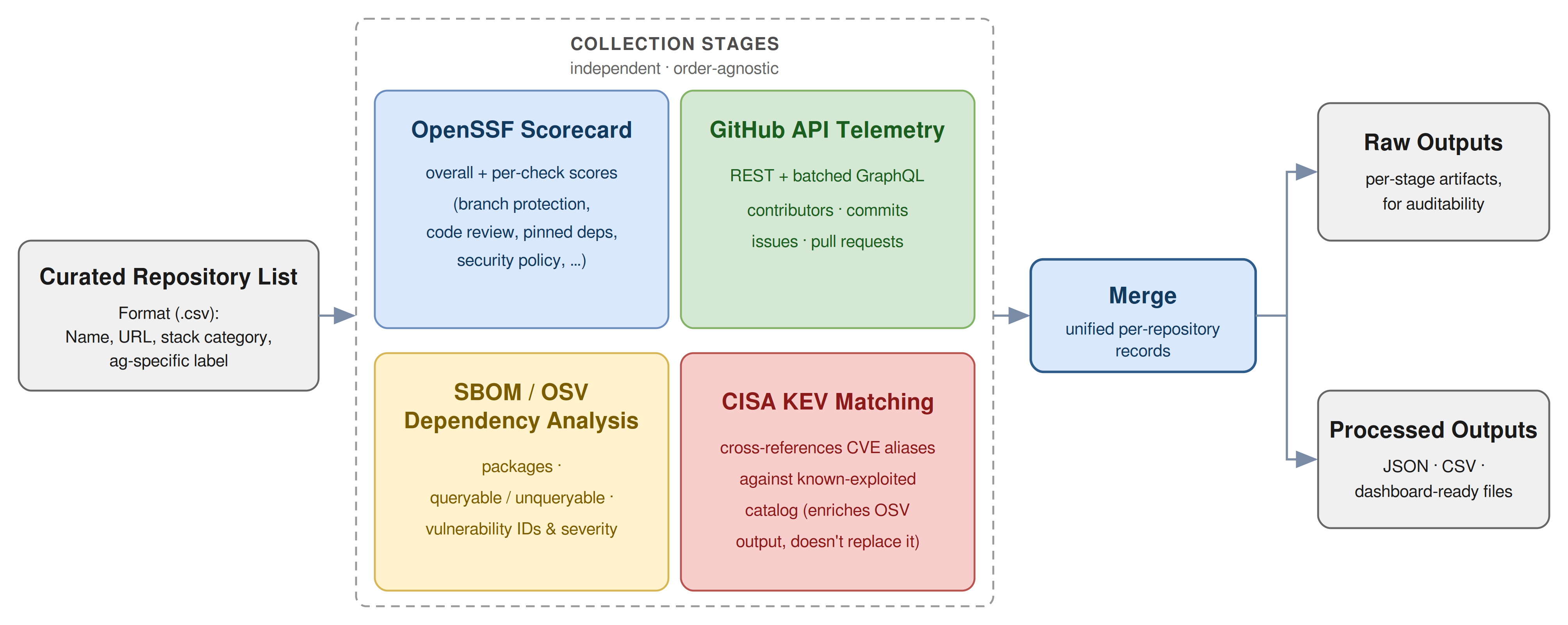}
    \caption{AgOSS data extraction pipeline. Four independent stages---OpenSSF Scorecard, GitHub API telemetry, SBOM resolution with OSV lookup, and CISA KEV matching---run per repository and merge into one analysis table.}
    \Description{Data Extraction Workflow diagram showcasing the input repository, the data collection stages (OpenSSF Scorecard, GitHub API, SBOM, CISA KEY), how they interact with one another, and the outputs (raw and processed) produced once the data has been collected.}
    \label{fig:pipeline}
\end{figure*}

We collected data via four collection stages: OpenSSF Scorecard, GitHub API telemetry, SBOM/OSV dependency analysis, and KEV matching, followed by a merge stage that is used for downstream data analyses (Figure~\ref{fig:pipeline}).

\paragraph{Metrics}
OpenSSF Scorecard is used as the repository-level security-posture signal~\cite{zahan2023scorecard}. It evaluates repositories against a set of supply-chain and repository-hygiene checks, each scored 0--10 (10 = strongest), and combines them into an overall score as a risk-weighted average. Development-activity and community-engagement telemetry is collected from the GitHub REST and GraphQL APIs, providing repository-level signals on contributor activity, commit activity, and issue/pull-request activity. Dependency exposure is measured via a separate SBOM/OSV stage. GitHub's dependency-graph endpoint is used to retrieve an SBOM (Software Bill of Materials)~\cite{githubsbom}. OSV is then used to identify known vulnerabilities affecting those dependencies. To separate general known vulnerabilities from those with stronger real-world evidence of exploitation, we additionally cross-reference OSV results against CISA's KEV catalog~\cite{cisa_kev}.

These sources yield ten core metrics: Scorecard overall, contributor count, commit count, open issues, closed issues, merged pull requests, star count, vulnerability count, vulnerability density (vulnerabilities per queryable package), and KEV-exploitable count. Together with the 18 individual Scorecard checks reported in Table~\ref{tab:scorecard-full}, these form the outcome set for all group comparisons.

\subsection{Statistical Tests}
\label{sec:stats}
Because the two groups are small and uneven (ag $n=55$, non-ag $n=11$), we report effect sizes as the primary evidence---rank-biserial $r$ for two-group comparisons, standardized mean difference for covariate balance---and treat $p$-values as secondary. All $p$-values are Benjamini--Hochberg (BH) corrected within their test family~\cite{benjamini1995controlling}.
 
For RQ1 we compare ag-specific against non-ag-specific repositories on each metric with the Mann--Whitney U test~\cite{mann1947test}, reporting rank-biserial $r$ with bootstrap 95\% confidence intervals, and compute Spearman's $\rho$ over all pairs of core metrics to test whether governance and dependency exposure move together. For RQ2 we compare the six stack categories with the Kruskal--Wallis test~\cite{kruskal1952use}, following significant omnibus results with Bonferroni-corrected Dunn post-hoc pairs~\cite{dunn1964multiple}.
 
RQ3 requires adjustment, since the univariate tests confound ag-specificity with category and size, and two categories (Domain-Specific Agricultural Platform and Field-Deployed Sensor) have no non-ag-specific members. We model the three outcomes with enough variance to support regression---Scorecard overall, vulnerability count, and vulnerability density---on \texttt{ag\_specific}, dummy-coded category with Cloud-hosted Backends and Dashboards as the reference, $\log(1+\text{stars})$, and $\log(1+\text{contributor count})$, using HC3 robust standard errors~\cite{mackinnon1985some}. We read this against the matched comparison of \S~\ref{sec:matched-methodology}, which fixes size and maturity by construction rather than by model specification. Within matched sets, we test the per-repository paired differences with the Wilcoxon signed-rank test~\cite{wilcoxon1945}, reporting the median difference with a nonparametric bootstrap 95\% confidence interval over 2,000 resamples.


\section{Results}
Scorecard ran on all 66 repositories (48 full, 18 partial but valid) and GitHub telemetry was available for all. Dependency resolution succeeded for 55 (44 ag-specific, 11 non-ag-specific); the 11 failures were all ag-specific repositories with no queryable SBOM, so exposure comparisons involving ag-specific software are conservative.
 
The dependency scan covered 20{,}050 packages (19{,}644 queryable) and returned 2{,}841 findings across 1{,}649 unique vulnerability IDs (167 Critical, 893 High, 818 Medium, 240 Low, 723 Unknown). Of the 55 scanned repositories, 36 (65.5\%) carried at least one vulnerability.
 
\subsection{RQ1: The Unadjusted Baseline Comparison}
\label{sec:mannwhitney}
 
Ag-specific repositories ($n=55$) had a median OpenSSF Scorecard of 3.3 (mean 3.4, 95\% CI [3.1, 3.7]); non-ag-specific repositories ($n=11$) reached 5.2 (mean 5.4, 95\% CI [5.0, 5.9]). Table~\ref{tab:scorecard-full} reports per-check results across all 66 repositories.
 
BH correction was applied across the 28 Mann--Whitney tests in this family (Table~\ref{tab:mannwhitney}). Non-agricultural repositories led on every activity, popularity, and governance metric, all large effects; the Scorecard gap was the widest ($r=-0.873$, $p_{\mathrm{FDR}}<0.001$). Dependency exposure was the exception: ag-specific repositories carried fewer raw vulnerabilities ($r=-0.424$, $p_{\mathrm{FDR}}=0.044$), matching their smaller dependency footprints, and neither vulnerability density (vulnerabilities per queryable package) nor KEV count differed. Among individual checks, the gap concentrated in CI, Maintained, Branch-Protection, and Security-Policy.
 
\begin{figure}[ht]
\centering
\includegraphics[width=\columnwidth]{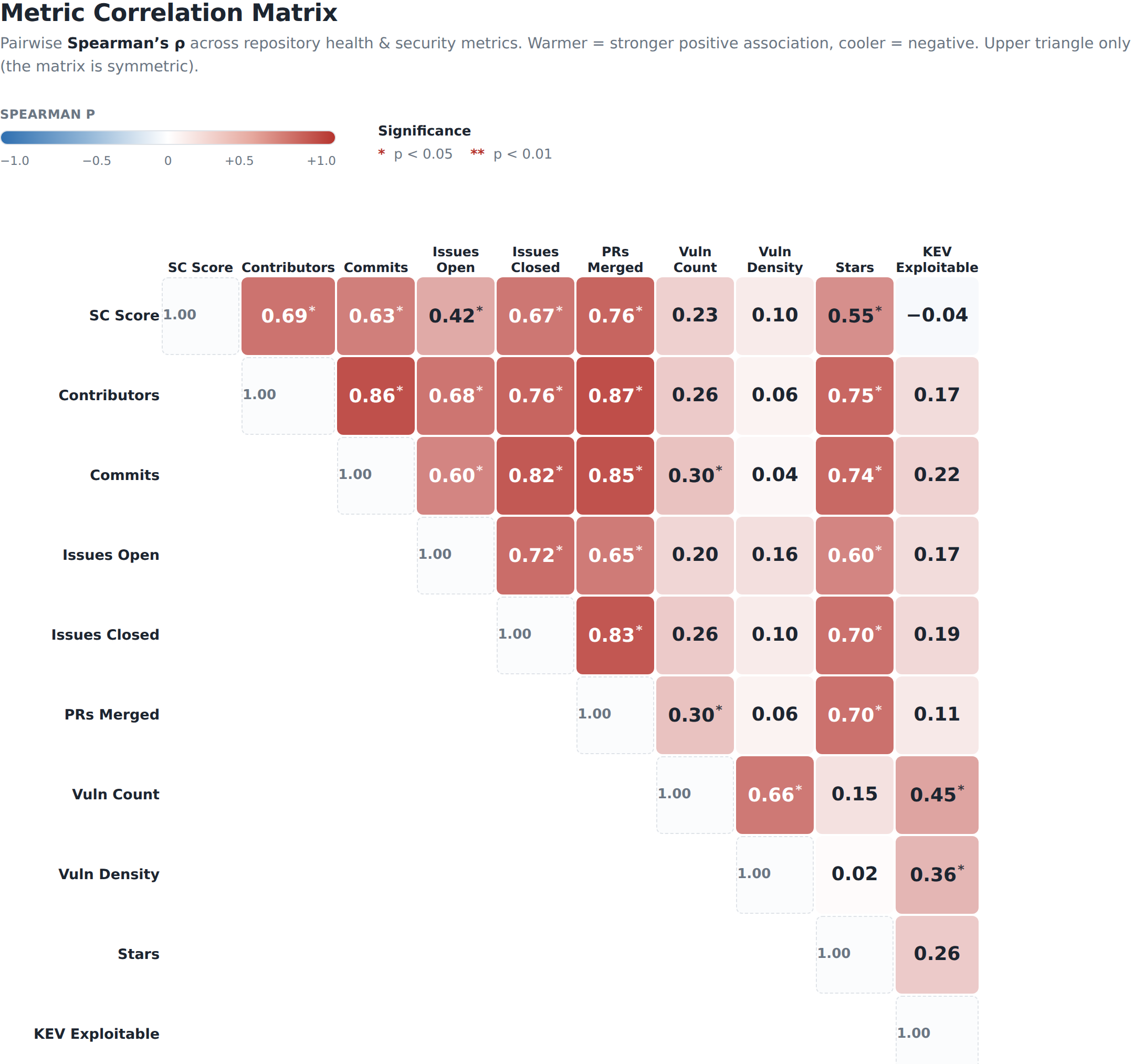}
\caption{Metric correlation matrix (Spearman $\rho$)}
\Description{Metric Correlation Matrix showcasing the BH-FDR corrected p-values.}
\label{fig:correlation-matrix}
\end{figure}

Spearman's $\rho$ over the $\binom{10}{2}=45$ metric pairs, BH-corrected once over the full set, is shown in Figure~\ref{fig:correlation-matrix}. Activity and community metrics form one tightly correlated cluster. Scorecard sits inside it but is decoupled from dependency risk: its correlations with vulnerability count ($\rho=0.23$, $p_{\mathrm{FDR}}=0.13$), density ($\rho=0.10$, $p_{\mathrm{FDR}}=0.53$), and KEV count ($\rho=-0.04$, $p_{\mathrm{FDR}}=0.82$) all fail correction, while the dependency-risk metrics correlate among themselves as expected (count with KEV, $\rho=0.45$; count with density, $\rho=0.66$; both significant).
 
Real-world exploitation reaches only a small subset of these vulnerabilities. Seven KEV-matched advisory records, resolving to 6 distinct CVEs (two records alias to the same CVE), affect 10 repositories (18.2\% of those scanned) across 5 of the 6 categories. Counted as repository–advisory instances, exposure concentrates in Edge and Gateway Software (3 repositories, 5 instances) and Cloud-hosted Backends and Dashboards (3 repositories, 4 instances); the remaining 4 affected repositories are spread across the other three categories. No repository in Data Processing Libraries/Tools carried a KEV match, but its small size ($n = 6$) makes that absence uninformative. The actively exploited weaknesses are thus spread across layers rather than confined to one, and this is a lower bound, since KEV lists only confirmed public exploitation.
 
However, this comparison is confounded by scale: the non-ag-specific repositories are large projects, averaging 15{,}548 stars and 983 contributors against 186 and 17 for the ag-specific set. \S~\ref{sec:matched-comparison} separates scale from domain using external matched controls.
 
\begin{table}[t]
\centering
\caption{OpenSSF Scorecard check results ($n=66$). N/E = not evaluable (absent/could not be assessed). Risk = OpenSSF severity (H/M/L/C). Full results in pipeline repository (see \textit{Data and Artifact Availability}).}
\label{tab:scorecard-full}
\footnotesize
\renewcommand{\arraystretch}{1.15}
\begin{tabularx}{\columnwidth}{@{}l >{\raggedright\arraybackslash}X rrr c@{}}
\toprule
\textbf{Category} & \textbf{Check} & \textbf{Pos.} & \textbf{Zero} & \textbf{N/E} & \textbf{Risk} \\
\midrule
Code Vln & Vulnerabilities & 39 & 27 & 0 & H \\
\midrule
\multirow{5}{*}{Maint.}
 & Dependency-Update-Tool & 17 & 49 & 0 & H \\
 & Maintained & 33 & 33 & 0 & H \\
 & Security-Policy & 11 & 55 & 0 & M \\
 & License & 60 & 6 & 0 & L \\
 & CII-Best-Practices & 3 & 63 & 0 & L \\
\midrule
\multirow{3}{*}{Cont.\ Test}
 & CI-Tests & 24 & 28 & 14 & L \\
 & Fuzzing & 3 & 63 & 0 & M \\
 & SAST & 7 & 59 & 0 & M \\
\midrule
\multirow{5}{*}{Source}
 & Binary-Artifacts & 58 & 8 & 0 & H \\
 & Branch-Protection & 15 & 33 & 18 & H \\
 & Dangerous-Workflow & 38 & 3 & 25 & C \\
 & Code-Review & 38 & 28 & 0 & H \\
 & Contributors & 55 & 11 & 0 & L \\
\midrule
\multirow{4}{*}{Build}
 & Pinned-Dependencies & 6 & 45 & 15 & M \\
 & Token-Permissions & 8 & 33 & 25 & H \\
 & Packaging & 17 & 0 & 49 & M \\
 & Signed-Releases & 0 & 21 & 45 & H \\
\bottomrule
\end{tabularx}
\end{table}
 
\begin{table*}[t]
\centering
\caption{Mann--Whitney U results, ag ($n_{\mathrm{ag}}$) vs.\ non-ag ($n_{\mathrm{non}}$), \textbf{FDR-significant metrics only}. $r$ is the rank-biserial effect size with 95\% bootstrap CI; $p_{\mathrm{FDR}}$ applies Benjamini--Hochberg correction across all 28 tests in the family. Negative $r$ indicates ag-specific repositories score lower. Ten other metrics showed no significant difference after correction and are omitted. Full 28-metric results are in the supplementary pipeline repository (see \textit{Data and Artifact Availability}).}
\label{tab:mannwhitney}
\footnotesize
\setlength{\tabcolsep}{6pt}
\begin{tabular}{@{}llrrrrrrl@{}}
\toprule
\textbf{Group} & \textbf{Metric} & $\bm{n_{\mathrm{ag}}}$ & $\bm{n_{\mathrm{non}}}$ &
$\bm{U}$ & $\bm{p}$ & $\bm{p_{\mathrm{FDR}}}$ & $\bm{r}$ & \textbf{95\% CI} \\
\midrule
\multirow{8}{*}{Repository}
 & Scorecard Overall   & 55 & 11 & 38.5  & $<$0.001 & $<$0.001 & $-$0.873 & [$-$0.970, $-$0.739] \\
 & Contributor Count   & 55 & 11 & 14.5  & $<$0.001 & $<$0.001 & $-$0.952 & [$-$1.000, $-$0.874] \\
 & Commit Count        & 55 & 11 & 69.0  & $<$0.001 & $<$0.001 & $-$0.772 & [$-$0.921, $-$0.570] \\
 & Issues Opened       & 55 & 11 & 74.0  & $<$0.001 & $<$0.001 & $-$0.755 & [$-$0.944, $-$0.514] \\
 & Issues Closed       & 55 & 11 & 33.0  & $<$0.001 & $<$0.001 & $-$0.891 & [$-$0.977, $-$0.765] \\
 & PRs Merged          & 55 & 11 & 47.0  & $<$0.001 & $<$0.001 & $-$0.845 & [$-$0.957, $-$0.696] \\
 & Stars Count         & 55 & 11 & 13.0  & $<$0.001 & $<$0.001 & $-$0.957 & [$-$1.000, $-$0.868] \\
 & Vulnerability Count & 44 & 11 & 139.5 & 0.028    & 0.044    & $-$0.424 & [$-$0.719, $-$0.087] \\
\midrule
\multirow{10}{*}{Scorecard Check}
 & CI-Tests               & 41 & 11 & 54.0  & $<$0.001 & $<$0.001 & $-$0.760 & [$-$0.911, $-$0.588] \\
 & Maintained             & 55 & 11 & 95.0  & $<$0.001 & $<$0.001 & $-$0.686 & [$-$0.830, $-$0.521] \\
 & Branch-Protection$^{\dagger}$ & 43 & 5 & 35.5 & 0.003 & 0.006 & $-$0.670 & [$-$0.967, $-$0.195] \\
 & Security-Policy        & 55 & 11 & 128.5 & $<$0.001 & $<$0.001 & $-$0.575 & [$-$0.855, $-$0.274] \\
 & Dependency-Update-Tool & 55 & 11 & 132.0 & $<$0.001 & 0.001    & $-$0.564 & [$-$0.800, $-$0.273] \\
 & Code-Review            & 55 & 11 & 165.0 & 0.014    & 0.024    & $-$0.455 & [$-$0.754, $-$0.098] \\
 & Contributors (check)   & 55 & 11 & 181.5 & 0.013    & 0.024    & $-$0.400 & [$-$0.527, $-$0.273] \\
 & Binary-Artifacts       & 55 & 11 & 405.5 & 0.019    & 0.031    & $+$0.341 & [$+$0.038, $+$0.636] \\
 & CII-Best-Practices     & 55 & 11 & 220.0 & $<$0.001 & $<$0.001 & $-$0.273 & [$-$0.545, 0.000] \\
 & Fuzzing                & 55 & 11 & 220.0 & $<$0.001 & $<$0.001 & $-$0.273 & [$-$0.545, 0.000] \\
\bottomrule
\end{tabular}
\par\vspace{2pt}
{\footnotesize All rows significant at $p_{\mathrm{FDR}}<0.05$. Effect magnitude ($|r|$):
$<$0.1 negligible, 0.1--0.3 small, 0.3--0.5 medium, $\geq$0.5 large.
$^{\dagger}$Non-ag group $n=5$; interpret with caution despite the large $r$.\par}
\end{table*}
 
\subsection{RQ2: Distribution Across Stack Layers}
BH correction was applied within two families: the 9 core-metric omnibus tests (star count was not run in this battery), which include Scorecard overall, and the individual-check omnibus tests. Of the 18 checks, 16 entered the correction; Packaging and Signed-Releases were constant across the corpus, with every scored repository returning 10 and 0 respectively, leaving the omnibus test undefined.
 
Median Scorecard scores ran from 4.9 (Domain-Specific Agricultural Platform, $n=9$) down to 2.7 (Field-Deployed Sensor, $n=14$), but the spread was not significant ($H=9.39$, $p_{\mathrm{FDR}}=0.142$), so no Dunn pair was read as a confirmed difference. Across the 9 core metrics, five were nominally significant but none survived correction (all $p_{\mathrm{FDR}}\geq0.071$). Only two checks survived: Security-Policy ($H=17.14$, $p_{\mathrm{FDR}}=0.034$) and the raw Vulnerabilities check ($H=20.57$, $p_{\mathrm{FDR}}=0.016$), the latter driven by Field-Deployed Sensor vs.\ Cloud-hosted Backends (Dunn $Z=3.61$, $p_{\mathrm{FDR}}=0.005$). Category membership is therefore a weak predictor of engineering maturity here; the two surviving checks warrant follow-up but rest on modest per-category samples ($n=5$ to $19$).
 
 Absolute exposure was uneven even where the omnibus was null (Table~\ref{tab:vuln-by-category}). Cloud-hosted Backends and Dashboards carried the most vulnerabilities overall (960 across 16 scanned repositories), while Embedded OS Substrate had the highest per-repository average (71.4). These averages are fragile, particularly within Domain-Specific Agricultural Platforms, where Ekylibre alone accounts for 287/553 instances (52\%), so per-category means describe individual projects as much as categories and we base no inferential claim on them.
 
\begin{table}[t]
\caption{Vulnerable dependencies by category (successfully scanned repositories, $n=55$).}
\label{tab:vuln-by-category}
\centering
\small
\setlength{\tabcolsep}{4pt}
\begin{tabularx}{\columnwidth}{>{\raggedright\arraybackslash}X r r r}
\toprule
\textbf{Category} & \textbf{Repos} & \textbf{Vulns} & \textbf{Avg./Repo} \\
\midrule
Cloud Backends \& Dashboards & 16 & 960 & 60.0 \\
Domain-Specific Ag.\ Platform & 8 & 553 & 69.1 \\
Edge and Gateway Software & 11 & 520 & 47.3 \\
Embedded OS Substrate & 5 & 357 & 71.4 \\
Field-Deployed Sensor & 9 & 431 & 47.9 \\
Data Processing Libraries/Tools & 6 & 20 & 3.3 \\
\bottomrule
\end{tabularx}
\end{table}

\subsection{RQ3: Separating Domain from Maturity}

\subsubsection{Joint Regression Models}
\label{sec:joint-models}
 
\begin{table}[t]
\caption{Joint OLS models: \texttt{ag\_specific} coefficient after adjusting for category, $\log(\mathrm{stars})$, and $\log(\mathrm{contributors})$. No outcome reached significance after FDR correction.}
\label{tab:joint-models}
\centering
\small
\begin{tabularx}{\columnwidth}{>{\raggedright\arraybackslash}Xccc}
\toprule
\textbf{Outcome} & \textbf{Coef.} & \textbf{95\% CI} & $\boldsymbol{p_{\mathrm{FDR}}}$ \\
\midrule
Scorecard Overall ($n=66$, $R^2=0.57$) &
$-$0.41 &
[$-$1.31, 0.49] &
0.431 \\

$\log(1+\mathrm{Vuln.\ Count})$ ($n=55$, $R^2=0.37$) &
$-$1.62 &
[$-$3.57, 0.33] &
0.313 \\

$\log(1+\mathrm{Vuln.\ Density})$ ($n=55$, $R^2=0.23$) &
$-$0.37 &
[$-$1.28, 0.55] &
0.431 \\
\bottomrule
\end{tabularx}
\end{table}
 
Once category, popularity, and community size are held constant, \texttt{ag\_specific} is not significant in any model ($p_{\mathrm{FDR}}=0.31$--$0.43$; Table~\ref{tab:joint-models}). Its confidence intervals cross zero before correction as well, so the null does not depend on the adjustment; category and size coefficients below are raw, uncorrected p-values. What survives is category; domain does not. On Scorecard, the only significant positive predictor is $\log(\text{contributor count})$ ($\beta=0.31$, $p=0.022$), while Data Processing Libraries/Tools ($\beta=-0.65$) and Field-Deployed Sensor ($\beta=-0.66$) score significantly lower; on log vulnerability count, Data Processing Libraries/Tools ($\beta=-2.86$) and Edge and Gateway Software ($\beta=-2.24$) carry significantly fewer. A collinearity caveat applies to reading these coefficients individually. VIFs were 3.3--3.4 for \texttt{ag\_specific} and 4.2--4.5 for $\log(\text{stars})$ and $\log(\text{contributors})$, all under the conventional threshold of 5. This does not bias the estimates, but it roughly doubles the standard errors on the two size covariates, so the shared scale signal is not split between stars and contributors---they are better read together than apart. It also costs precision on \texttt{ag\_specific}, making its null a failure to detect a domain effect at this sample size rather than a demonstration that none exists. The matched analysis in \S~\ref{sec:matched-comparison} approaches the same question by fixing size and maturity through design rather than model specification, and so does not inherit this collinearity.
 
\subsubsection{Matched Comparisons}
\label{sec:matched-comparison}
47 of the 55 ag-specific repositories (85.5\%) matched under the primary caliper; 1 was dropped for having no control sharing its primary language (Julia) and 7 for having no control inside the distance caliper (mean nearest-neighbor Mahalanobis distance 1.35, median 1.31). The 8 unmatched repositories were ADAPT/ADMPlugin, ADAPT/StandardPlugin, APSIMInitiative/ApsimX, Digital-Naturalism-Laboratories/Mothbox, LiteFarmOrg/LiteFarm, Sen2Agri/Sen2Agri-System, agstack/weather-server, and autogrow/openminder.
 
\begin{figure}[ht]
\centering
\includegraphics[width=\columnwidth]{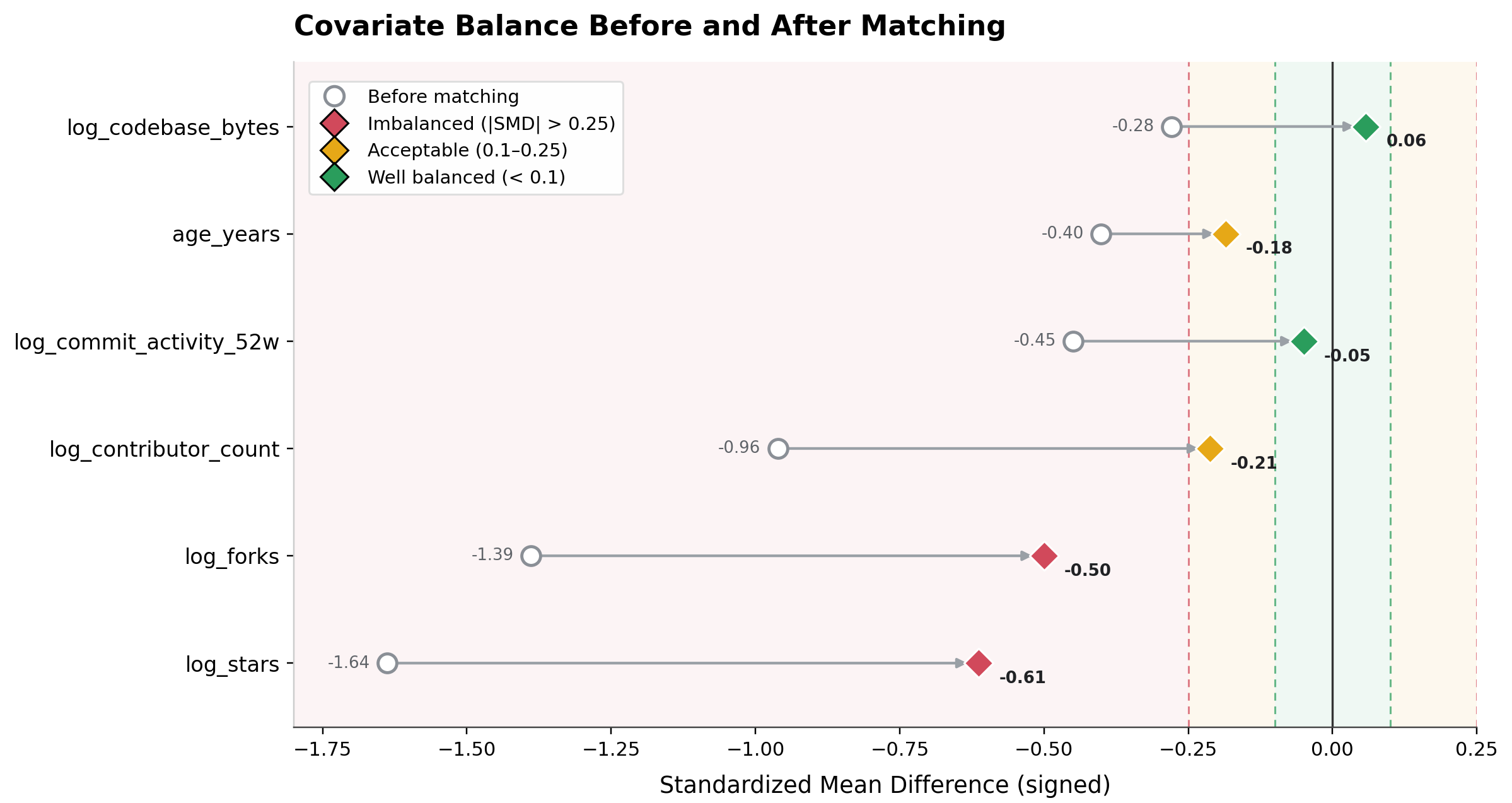}
\caption{Standardized mean difference (SMD) of matching covariates before vs.\ after matching ($k=3$).}
\Description{Comparing the SMD of matching covariates before and after the matched comparison analysis is done}
\label{fig:matched-balance}
\end{figure}
 
Matching improved covariate balance without fully closing it (Figure~\ref{fig:matched-balance})~\cite{austin2011introduction}. SMD on log commit activity fell from $-0.45$ to $-0.05$ and on log codebase size from $-0.28$ to $+0.06$ (both well balanced), while age ($-0.40$ to $-0.18$) and log contributor count ($-0.96$ to $-0.21$) became acceptable. Log stars ($-1.64$ to $-0.61$) and log forks ($-1.39$ to $-0.50$) improved but stayed above the imbalance threshold, which motivates the regression-adjusted estimate below.
 
\begin{table}[t]
\caption{Matched-pair comparison ($k=3$); no comparison reached
significance after FDR correction.}
\label{tab:matched-headline}
\centering
\small
\setlength{\tabcolsep}{3pt}
\begin{tabularx}{\columnwidth}{>{\raggedright\arraybackslash}X c c c c}
\toprule
\textbf{Metric} &
\shortstack{\textbf{Median $\Delta$}\\[-2pt] \footnotesize (repo $-$ ctrl)} &
\textbf{95\% CI} &
$\boldsymbol{p_{\mathrm{FDR}}}$ &
$\boldsymbol{n}$ \\
\midrule
Scorecard Overall     & $-$0.40 & [$-$0.50, $-$0.10] & 0.170 & 47 \\
Vulnerability Count   & $-$8.75 & [$-$25.00, 0.00]   & 0.389 & 38 \\
Vulnerability Density & 0.00    & [$-$0.04, 0.00]    & 0.820 & 38 \\
KEV Exploitable Count & 0.00    & [0.00, 0.00]       & 0.397 & 47 \\
\bottomrule
\end{tabularx}
\end{table}
 
Against matched controls, the raw Scorecard gap collapses. The matched median difference is $-0.40$ ($r=-0.362$, $p_{\mathrm{FDR}}=0.170$), down from $r=-0.873$ unmatched, and vulnerability count, density, KEV count all sit at or near zero (Table~\ref{tab:matched-headline}). The residual $r = -0.362$ is a medium effect (Table~\ref{tab:mannwhitney}), so $p_{FDR} = 0.170$ at $n = 47$ reflects limited power, not established absence. Notably, the covariates that remain imbalanced after matching (log stars, log forks; Fig.~\ref{fig:matched-balance}) leave controls more popular than their ag-specific matches, so the imbalance runs against the ag-specific repositories but the gap still collapses.
 
Under the random-matching robustness check, median Scorecard differences were $-0.30$ at $k=1$, $-0.43$ at $k=3$, and $-0.48$ at $k=5$, negative throughout and never significant. Across 1{,}000 random-matching seeds at $k=3$, the Scorecard direction was stable in every seed (seed-to-seed SD $0.11$), whereas dependency-update-tool (67\% of seeds agreeing on direction), maintained (57\%), vulnerability count (71\%), and vulnerability density (76\%) were seed-dependent. Bounding the 8 unmatched repositories (mentioned by name earlier) against their best- and worst-case observed outcomes left Scorecard as the only direction-stable metric (best $-0.27$, worst $-0.47$, both negative) whereas every other metric changed direction under at least one bound. The lone surviving direction, a lower Scorecard, did not hold up to regression adjustment (Table~\ref{tab:matched-regression}). Fitting each matched outcome on treatment plus the two still-imbalanced covariates ($\log$ stars, $\log$ forks) moved the Scorecard estimate from $-0.31$ [$-0.57$, $-0.05$] to $+0.06$ [$-0.31$, $0.44$]. No outcome was significant under both the matching-only and regression-adjusted estimates.
 
\begin{table}[t]
\caption{Matching-only vs.\ matching + regression estimates on the
matched sample (47 treated repositories, 83 distinct controls;
adjusting for $\log$ stars and $\log$ forks). Estimates are means;
cf.\ the medians in Table~\ref{tab:matched-headline}.}
\label{tab:matched-regression}
\centering
\footnotesize
\setlength{\tabcolsep}{4pt}
\renewcommand{\arraystretch}{1.15}
\begin{tabularx}{\columnwidth}{
  >{\raggedright\arraybackslash}X
  >{\centering\arraybackslash}p{0.15\columnwidth}
  >{\centering\arraybackslash}p{0.15\columnwidth}
  >{\centering\arraybackslash}p{0.11\columnwidth}}
\toprule
\textbf{Outcome} &
\textbf{Match-only mean $\Delta$} &
\textbf{Match+Reg.\ $\beta$} &
\textbf{Same dir.?} \\
\midrule
Scorecard Overall & $-$0.31 & $+$0.06 & No \\
Vulnerability Count & $-$1.23 & $+$7.05 & No \\
Vulnerabilities (Scorecard check) & $-$1.34 & $-$1.62 & Yes \\
Dependency-Update-Tool & $-$1.42 & $-$0.92 & Yes \\
\bottomrule
\end{tabularx}
\end{table}
 

\section{Analysis and Discussion}

\subsection{A Confounded Maturity Gap}
\label{sec:discussion-maturity-gap}
The unadjusted comparison in \S~\ref{sec:mannwhitney} points to a domain deficit: non-ag-specific repositories lead on Scorecard ($r=-0.873$) and on every activity and popularity metric ($|r|>0.75$). This echoes a concern raised for the domain: dos Santos \textit{et al.}~\cite{jsan12020028} found weak repository management across precision-agriculture tools, and the smart-farming security literature treats agriculture as a risk surface~\cite{barreto2018survey,gupta2020security}. Both of our adjustment strategies undercut that reading. Joint regression (\S~\ref{sec:joint-models}) leaves \texttt{ag\_specific} non-significant for all three outcomes ($p_{\mathrm{FDR}}=0.31$--$0.43$) while contributor count stays predictive; matching (\S~\ref{sec:matched-comparison}) shrinks the Scorecard gap from $r=-0.873$ to $r=-0.362$, which no longer survives correction. Therefore, we can no longer distinguish an agricultural repository from a non-agricultural project of similar language, size, age, and activity on governance posture. This matches what Zahan \textit{et al.} found in npm and PyPI, where raw practice-outcome associations were confounded by project characteristics and cleared only after matching or regression~\cite{zahan2023practices,zahan2025prioritizing}. 

Contributor-community size is the variable most consistently tied to maturity: the strongest non-activity correlate of Scorecard ($\rho=0.69$), the only significant positive predictor in the joint Scorecard model ($\beta=0.31$, $p=0.022$), and the sharpest single separator of the two groups in the raw comparison ($r=-0.952$). Zahan \textit{et al.}~\cite{zahan2022weaklinks} also rank too-few maintainers among npm's strongest weak links, and Zimmermann \textit{et al.}~\cite{zimmermann} show how a handful of maintainers become implicitly trusted across the ecosystem. The practical implication is that the projects struggling on governance are the thinly-staffed ones, and contributor capacity is the lever rather than ag-specificity.

\subsection{Supply-Chain Exposure Beyond Governance}
\label{sec:independent-repo-gov}

Scorecard is decoupled from vulnerability count, density, and KEV count, and the raw comparison splits the same way, separating the groups on every governance metric but not on density or KEV count. A repository's Scorecard says little about the dependency risk it carries---the same disconnect that Zahan \textit{et al.}~\cite{zahan2023practices} reported in npm and PyPI, where higher scores sometimes tracked slightly more vulnerabilities. The reason is structural. Scorecard measures what a maintainer controls directly while vulnerabilities arrive with the upstream packages a project pulls in~\cite{decan2018impact,prana2021outofsight}. Zimmermann \textit{et al.}~\cite{zimmermann} show that npm risk concentrates in the reach and shape of the dependency graph, with vulnerable code persisting downstream long after a patch ships. A disciplined project with a wide transitive graph can inherit heavy exposure; a loosely-run project with few dependencies can carry almost none. 

The joint models place the signal in category, not domain, in both outcomes. On vulnerability count, Data Processing Libraries / Tools and Edge and Gateway Software fall significantly below the Cloud-hosted Backends reference; on Scorecard, Data Processing Libraries/Tools and Field-Deployed Sensor do. The \texttt{ag\_specific} term carries neither. Part of that spread is genuine variation in dependency-graph breadth, and part is how completely GitHub resolves each ecosystem---better for npm and PyPI than for the C and embedded build systems in the lower layers~\cite{balliu2023challenges,imtiaz2021comparative}. On this evidence, raising Scorecard checks would not lower inherited exposure---the two dimensions need separate measurement and separate remediation.

\subsection{Representative Repository Case Studies}
\label{sec:case-studies}
To ground these aggregate results, we examined five representative repositories from different parts of the AgOSS dataset in Table~\ref{tab:case-studies}.

\begin{table}[t]
\centering
\caption{Repository case studies.}
\label{tab:case-studies}
\footnotesize
\setlength{\tabcolsep}{4pt}
\renewcommand{\arraystretch}{1.15}
\begin{tabular}{@{}p{0.53\linewidth} p{0.44\linewidth}@{}}
\toprule
\textbf{Repository} & \textbf{Key finding} \\
\midrule
\textbf{OpenATK FieldWorkApp} --- Ag $\cdot$ JS $\cdot$ 8.4\,yrs $\cdot$ 3 contrib., 1 star. Scorecard \textbf{1.5} (lowest); 1,741 deps $\rightarrow$ 235 vulns, incl.\ KEV CVE-2023-5217.
& A small team can inherit an actively exploited attack surface through dependency choices alone. \\
\addlinespace[2pt]
\textbf{FarmOS} --- Ag $\cdot$ PHP $\cdot$ 12.4\,yrs $\cdot$ 54 contrib., 1,308 stars. Scorecard \textbf{5.7} (highest ag), above non-ag median 5.2. 48 deps $\rightarrow$ 6 vulns.
& The ag vs.\ non-ag gap (\S~\ref{sec:mannwhitney}) is a tendency, not a rule. \\
\addlinespace[2pt]
\textbf{NodeODM} --- Non-ag $\cdot$ JS $\cdot$ 9.9\,yrs $\cdot$ 31 contrib., 294 stars. Scorecard \textbf{4.4} (lowest non-ag, above ag median 3.3); 41 deps $\rightarrow$ 11 vulns, incl.\ KEV CVE-2021-21315.
& The complement of FarmOS: non-ag status confers no immunity to weak practices or exploit exposure. \\
\addlinespace[2pt]
\textbf{Zephyr} --- Non-ag $\cdot$ C $\cdot$ 10.1\,yrs $\cdot$ 4,361 contrib., 15,850 stars. Scorecard \textbf{7.1} (highest); 675 deps $\rightarrow$ 229 vulns, incl.\ KEV CVE-2023-4863.
& Strong governance cannot prevent inheriting an exploited dependency footprint in a foundational substrate (\S~\ref{sec:independent-repo-gov}). \\
\addlinespace[2pt]
\textbf{Ekylibre} --- Ag $\cdot$ Ruby $\cdot$ 14.9\,yrs $\cdot$ 76 contrib., 484 stars. Scorecard \textbf{5.1}; 1,687 deps $\rightarrow$ \textbf{287} vulns, $>\!\tfrac{1}{2}$ of its category's 553.
& Governance/risk independence (\S~\ref{sec:independent-repo-gov}) holds within populations too, and category averages are skewed by one project. \\
\bottomrule
\end{tabular}
\end{table}


\section{Threats to Validity}
\label{sec:threats}

{\large \textbf{(1) Construct Validity}}
\paragraph{Dataset Construction}
Keyword-based retrieval is limited by vocabulary mismatch and may fail to capture relevant repositories lacking agricultural terminology. Manual screening and snowball sampling may skew coverage toward well-documented or highly connected repositories. Rapid triage was performed by a single rater; although inclusion-biased, we did not measure its false-negative rate against a second screen.

\paragraph{License Heterogeneity}
Our corpus characterizes available agricultural software broadly rather than a strictly OSI-defined population (4 repositories distributed under source-available or non-OSI-approved licenses, 6 with no license file). We retained these because they occupy operational roles within the agricultural stack. 

\paragraph{Vulnerability-to-Exploit Gap}
Our methodology identifies known vulnerabilities through OSV lookups and exploitable ones via KEV matching, but a vulnerability does not necessarily translate to an immediate threat. The KEV catalog represents only a subset of exploitable vulnerabilities and should therefore be interpreted as a conservative lower bound on exploitability.

\vspace{0.1cm}
\noindent
{\large \textbf{(2) Internal Validity}}
\paragraph{Omission of Formal Bus Factor Calculations}
While we collect raw contributor telemetry and observe thinly-staffed projects, we do not perform the algorithmic inference required to estimate a formal bus factor. This limits our ability to directly compare the human-resource resilience of the agricultural open source software ecosystem against other software domains.

\paragraph{Residual Confounding}
Factors such as funding model, institutional backing, governance structure, or deployment context were not used in either adjustment strategy (matched comparison, joint regression model) and may therefore contribute to the remaining observed differences between groups~\cite{stuart2010matching,austin2011introduction}.

\vspace{0.1cm}
\noindent
{\large \textbf{(3) External Validity}}

While we curated a diverse dataset across different layers of the agricultural stack, our sample of 66 repositories represents a subset of the agricultural open source software ecosystem. The reliance on GitHub-centric discovery (stars, forks, and metadata) may overlook significant agricultural software hosted on alternative platforms like GitLab or Bitbucket instances used in industrial contexts. Our corpus is a purposive sample rather than a random draw from a defined super-population. The reported p-values should therefore be read as descriptive summaries of the observed groups rather than as inferences to all agricultural OSS.

\vspace{0.1cm}
\noindent
{\large \textbf{(4) Data and Conclusion Validity}}

\paragraph{SBOM Availability:} Eleven repositories (16.67\%) failed to return a valid SBOM via the GitHub endpoint (HTTP 404), preventing their inclusion in the dependency analysis. SBOM completeness varies sharply by ecosystem and tooling~\cite{balliu2023challenges, imtiaz2021comparative}, and our failures concentrate in C/C++ and embedded build systems (CMake, Kconfig, West), which GitHub's dependency graph resolves far less completely than manifest-based ecosystems (npm, PyPI). The Field-Deployed Sensor category is disproportionately affected: 5 of its 14 repositories (36\%) returned no SBOM, against 11 of 66 (17\%) corpus-wide.

\paragraph{Package Queryability:} Out of 20,050 package records, 406 were unqueryable due to non-standard naming conventions or missing version metadata, potentially masking additional risks.


\section{Conclusion}
We introduced AgOSS, a dataset of 66 OSS repositories spanning six categories, and measured their supply chain security through OpenSSF Scorecard, governance and activity metrics, SBOM-based dependency analysis, and KEV matching. Agricultural repositories score below general-purpose software on governance and maintenance, but this gap is confounded by scale rather than being intrinsic to the domain. It does not survive matching against size- and activity-comparable controls, nor size-adjusted regression, though at our relatively small sample size that is a failure to detect a domain difference rather than proof of none. Contributor-community size is the variable that most consistently tracks governance maturity. Dependency exposure moves independently of Scorecard, concentrating in a project's dependency footprint rather than its practices.

For a fragmented ecosystem underpinning food production, these results direct remediation effort (\textit{e.g.}, funding) toward contributor sustainability and dependency management. The field-deployed layer, where GitHub's dependency graph resolves least completely, is also the layer whose exposure we can least trust, so a low observed count there may reflect a smaller dependency surface or one we failed to observe; closing that requires SBOM tooling for C and embedded build systems. More broadly, the confounding we found is unlikely to be specific to agriculture: any domain-level comparison omitting maturity controls risks the same error.


\begin{acks}
The authors used generative AI tools to support multiple aspects of this research. ChatGPT (OpenAI) and Gemini (Google) were used to assist with writing refinement and prose revision. ChatGPT (OpenAI) was additionally used to assist with LaTeX formatting and typesetting. Claude (Anthropic, claude.ai) and Gemini (Google) were used to provide feedback and revision suggestions during manuscript development. Claude (Anthropic, claude.ai) was additionally used to support the strengthening and articulation of analysis by presenting it with study results for interpretive feedback; all analytical conclusions were independently verified and are the authors' own. Claude Code (Anthropic) was used to assist in developing data collection and analysis scripts for the MSR-based study, including SBOM and API-based analyses; all code was reviewed and validated by the authors. The authors take full responsibility for the accuracy and integrity of all content presented in this paper.
\end{acks}


\section*{Ethics and Privacy Statement}
This study did not involve human subjects. Our analysis is limited to aggregate architectural layers, statistical comparisons, and dependency-count summaries rather than actionable exploitation guidance. All named vulnerabilities are pre-existing public CISA KEV entries, so no disclosure was warranted. Finally, our inclusion of source-available and custom-licensed projects reflects the heterogeneous licensing practices of the agricultural OSS ecosystem rather than an endorsement of any particular licensing model, and respects the range of data-sovereignty and community-governance norms across the agricultural communities represented in the dataset.


\section*{Data and Artifact Availability}
The AgOSS dataset is available at \url{https://anonymous.4open.science/r/agoss-dataset-82AD/}.
The extraction pipeline and statistical analysis code are also available at \url{https://anonymous.4open.science/r/AgOSS-Data-Gatherer-D6E5/}

\bibliographystyle{ACM-Reference-Format}
\bibliography{references}

@online{grandview2026,
  author       = {{Grand View Research}},
  title        = {Precision Farming Market Size And Share Report, 2026--2033},
  year         = {2026},
  url          = {https://www.grandviewresearch.com/industry-analysis/precision-farming-market},
}

@techreport{ossra2026,
  author = {{Black Duck}},
  title = {Open Source Security and Risk Analysis Report},
  institution = {Black Duck Software},
  year = {2026},
  url = {https://www.blackduck.com/resources/analyst-reports/open-source-security-risk-analysis.html}
}

@techreport{mcfadden2023precision,
  author      = {McFadden, Jonathan and Njuki, Eric and Griffin, Terry},
  title       = {Precision Agriculture in the Digital Era: Recent Adoption on {U.S.} Farms},
  institution = {U.S. Department of Agriculture, Economic Research Service},
  number      = {EIB-248},
  year        = {2023},
  url         = {https://www.ers.usda.gov/publications/pub-details?pubid=105893}
}

@article{lowder2016,
  author  = {Lowder, Sarah K. and Skoet, Jakob and Raney, Terri},
  title   = {The Number, Size, and Distribution of Farms, Smallholder Farms,
             and Family Farms Worldwide},
  journal = {World Development},
  volume  = {87},
  pages   = {16--29},
  year    = {2016},
  issn    = {0305-750X},
  doi     = {10.1016/j.worlddev.2015.10.041}
}

@article{chandra2021digital,
  author  = {Chandra, Ranveer and Collis, Stewart},
  title   = {Digital Agriculture for Small-Scale Producers: Challenges and
             Opportunities},
  journal = {Communications of the ACM},
  volume  = {64},
  number  = {12},
  pages   = {75--84},
  year    = {2021},
  doi     = {10.1145/3454008}
}

@online{fao_farmos,
  author       = {{Food and Agriculture Organization of the United Nations}},
  title        = {{FarmOS}: Free and Open Source Farm Management Software
                  Platform and Development Community},
  howpublished = {FAO Agroecology Knowledge Hub},
  year         = {2021},
  url          = {https://www.fao.org/agroecology/in-action/detail/farmos-free-and-open-source-farm-management-software-platform-and-development-community/en},
}

@online{agl,
  author       = {{The Linux Foundation}},
  title        = {Automotive Grade {Linux}},
  howpublished = {Project website},
  year         = {2024},
  url          = {https://www.automotivelinux.org/},
}

@inproceedings{barreto2018survey,
  author    = {Barreto, Luiz and Amaral, Antonio},
  title     = {Smart Farming: Cyber Security Challenges},
  booktitle = {2018 International Conference on Intelligent Systems (IS)},
  pages     = {870--876},
  year      = {2018},
  publisher = {IEEE},
  doi       = {10.1109/IS.2018.8710531}
}

@article{gupta2020security,
  author  = {Gupta, Maanak and Abdelsalam, Mahmoud and Khorsandroo, Sajad and
             Mittal, Sudip},
  title   = {Security and Privacy in Smart Farming: Challenges and
             Opportunities},
  journal = {IEEE Access},
  volume  = {8},
  pages   = {34564--34584},
  year    = {2020},
  doi     = {10.1109/ACCESS.2020.2975142}
}

@article{jsan12020028,
  author    = {dos Santos, Rog{\'e}rio P. and Fachada, Nuno and Beko, Marko and
               Leithardt, Valderi R. Q.},
  title     = {A Rapid Review on the Use of Free and Open Source Technologies
               and Software Applied to Precision Agriculture Practices},
  journal   = {Journal of Sensor and Actuator Networks},
  volume    = {12},
  number    = {2},
  articleno = {28},
  numpages  = {18},
  year      = {2023},
  issn      = {2224-2708},
  doi       = {10.3390/jsan12020028}
}

@article{holzworth2015,
  author  = {Holzworth, Dean P. and Snow, Val and Janssen, Sander and
             Athanasiadis, Ioannis N. and Donatelli, Marcello and
             Hoogenboom, Gerrit and White, Jeffrey W. and Thorburn, Peter},
  title   = {Agricultural Production Systems Modelling and Software: Current
             Status and Future Prospects},
  journal = {Environmental Modelling \& Software},
  volume  = {72},
  pages   = {276--286},
  year    = {2015},
  doi     = {10.1016/j.envsoft.2014.12.013}
}

@preprint{tutko2022,
  author        = {Tutko, Adam and Henley, Austin Z. and Mockus, Audris},
  title         = {How are Software Repositories Mined? {A} Systematic
                   Literature Review of Workflows, Methodologies,
                   Reproducibility, and Tools},
  year          = {2022},
  archivePrefix = {arXiv},
  eprint        = {2204.08108},
  primaryClass  = {cs.SE},
  doi           = {10.48550/arXiv.2204.08108}
}

@inproceedings{githubfindings,
  author    = {Cosentino, Valerio and C{\'a}novas Izquierdo, Javier Luis and
               Cabot, Jordi},
  title     = {Findings from {GitHub}: Methods, Datasets and Limitations},
  booktitle = {Proceedings of the 13th International Conference on Mining
               Software Repositories (MSR)},
  pages     = {137--141},
  year      = {2016},
  publisher = {ACM},
  isbn      = {9781450341868},
  doi       = {10.1145/2901739.2901776}
}

@article{BaltesSebastian2022,
  author    = {Baltes, Sebastian and Ralph, Paul},
  title     = {Sampling in Software Engineering Research: A Critical Review and
               Guidelines},
  journal   = {Empirical Software Engineering},
  volume    = {27},
  number    = {4},
  articleno = {94},
  numpages  = {31},
  year      = {2022},
  issn      = {1382-3256},
  publisher = {Springer},
  doi       = {10.1007/s10664-021-10072-8}
}

@article{soliman2025,
  author    = {Soliman, Mohamed and Albonico, Michel and Malavolta, Ivano and Wortmann, Andreas},
  title     = {Mining Software Repositories for Software Architecture --- A Systematic Mapping Study},
  journal   = {Information and Software Technology},
  volume    = {181},
  pages     = {107677},
  year      = {2025},
  publisher = {Elsevier},
  doi       = {10.1016/j.infsof.2025.107677}
}

@article{prisma,
  author  = {Page, Matthew J and McKenzie, Joanne E and Bossuyt, Patrick M and
             Boutron, Isabelle and Hoffmann, Tammy C and Mulrow, Cynthia D and
             Shamseer, Larissa and Tetzlaff, Jennifer M and Akl, Elie A and
             Brennan, Sue E and Chou, Roger and Glanville, Julie and
             Grimshaw, Jeremy M and Hr{\'o}bjartsson, Asbj{\o}rn and
             Lalu, Manoj M and Li, Tianjing and Loder, Elizabeth W and
             Mayo-Wilson, Evan and McDonald, Steve and McGuinness, Luke A and
             Stewart, Lesley A and Thomas, James and Tricco, Andrea C and
             Welch, Vivian A and Whiting, Penny and Moher, David},
  title   = {The {PRISMA} 2020 Statement: An Updated Guideline for Reporting Systematic Reviews},
  journal = {BMJ},
  volume  = {372},
  pages   = {n71},
  year    = {2021},
  publisher = {BMJ Publishing Group},
  doi     = {10.1136/bmj.n71}
}

@online{hf2026incident,
  author = {{Hugging Face}},
  title = {Security incident disclosure - July 2026},
  year = {2026},
  url = {https://huggingface.co/blog/security-incident-july-2026},
}

@inproceedings{jiang2022,
  author    = {Jiang, Wenxin and Synovic, Nicholas and Sethi, Rohan and
               Indarapu, Aryan and Hyatt, Matt and Schorlemmer, Taylor R. and
               Thiruvathukal, George K. and Davis, James C.},
  title     = {An Empirical Study of Artifacts and Security Risks in the
               Pre-trained Model Supply Chain},
  booktitle = {Proceedings of the 2022 ACM Workshop on Software Supply Chain
               Offensive Research and Ecosystem Defenses (SCORED'22)},
  pages     = {105--114},
  numpages  = {10},
  year      = {2022},
  publisher = {Association for Computing Machinery},
  isbn      = {9781450398855},
  doi       = {10.1145/3560835.3564547}
}

@article{yu2026security,
  author  = {Yu, Hanbo and Khan, Faiyaz and Ding, Steven H. H. and Wu, Junjie
             and Stakhanova, Natalia and Fung, Benjamin C. M.},
  title   = {Security Risk Assessment of {Android} Automotive {OS} Software
             Supply Chain Using Firmware Reverse Engineering},
  journal = {Computers \& Security},
  volume  = {167},
  pages   = {104923},
  year    = {2026},
  issn    = {0167-4048},
  doi     = {10.1016/j.cose.2026.104923}
}

@inproceedings{zimmermann,
  author    = {Zimmermann, Markus and Staicu, Cristian-Alexandru and
               Tenny, Cam and Pradel, Michael},
  title     = {Small World with High Risks: A Study of Security Threats in the
               {npm} Ecosystem},
  booktitle = {Proceedings of the 28th USENIX Security Symposium
               (USENIX Security 19)},
  pages     = {995--1010},
  year      = {2019},
  publisher = {USENIX Association},
  isbn      = {978-1-939133-06-9},
  month     = aug,
  url       = {https://www.usenix.org/conference/usenixsecurity19/presentation/zimmerman}
}

@article{zahan2023scorecard,
  author    = {Zahan, Nusrat and Kanakiya, Parth and Hambleton, Brian and
               Shohan, Shohanuzzaman and Williams, Laurie},
  title     = {{OpenSSF} Scorecard: On the Path Toward Ecosystem-Wide Automated
               Security Metrics},
  journal   = {IEEE Security \& Privacy},
  volume    = {21},
  number    = {6},
  pages     = {76--88},
  year      = {2023},
  month     = nov,
  issn      = {1558-4046},
  publisher = {IEEE},
  doi       = {10.1109/MSEC.2023.3279773}
}

@inproceedings{zahan2023practices,
  author    = {Zahan, Nusrat and Shohan, Shohanuzzaman and Harris, Dan and
               Williams, Laurie},
  title     = {Do Software Security Practices Yield Fewer Vulnerabilities?},
  booktitle = {2023 IEEE/ACM 45th International Conference on Software
               Engineering: Software Engineering in Practice (ICSE-SEIP)},
  pages     = {292--303},
  year      = {2023},
  publisher = {IEEE},
  address   = {Melbourne, Australia},
  doi       = {10.1109/ICSE-SEIP58684.2023.00032}
}

@preprint{zahan2025prioritizing,
  author        = {Zahan, Nusrat and Williams, Laurie},
  title         = {Prioritizing Security Practice Adoption: Empirical Insights
                   on Software Security Outcomes in the {npm} Ecosystem},
  year          = {2025},
  archivePrefix = {arXiv},
  eprint        = {2504.14026},
  primaryClass  = {cs.CR},
  doi           = {10.48550/arXiv.2504.14026}
}

@inproceedings{zahan2022weaklinks,
  author    = {Zahan, Nusrat and Zimmermann, Thomas and Godefroid, Patrice and
               Murphy, Brendan and Maddila, Chandra and Williams, Laurie},
  title     = {What are Weak Links in the {npm} Supply Chain?},
  booktitle = {2022 IEEE/ACM 44th International Conference on Software
               Engineering: Software Engineering in Practice (ICSE-SEIP)},
  pages     = {331--340},
  year      = {2022},
  publisher = {IEEE},
  address   = {Pittsburgh, PA, USA},
  doi       = {10.1145/3510457.3513044}
}

@inproceedings{decan2018impact,
  author    = {Decan, Alexandre and Mens, Tom and Constantinou, Eleni},
  title     = {On the Impact of Security Vulnerabilities in the {npm} Package
               Dependency Network},
  booktitle = {Proceedings of the 15th International Conference on Mining
               Software Repositories (MSR)},
  pages     = {181--191},
  year      = {2018},
  publisher = {ACM},
  address   = {Gothenburg, Sweden},
  doi       = {10.1145/3196398.3196401}
}

@article{prana2021outofsight,
  author  = {Prana, Gede Artha Azriadi and Sharma, Abhishek and
             Shar, Lwin Khin and Foo, Darius and Santosa, Andrew E. and
             Sharma, Asankhaya and Lo, David},
  title   = {Out of Sight, Out of Mind? {How} Vulnerable Dependencies Affect
             Open-Source Projects},
  journal = {Empirical Software Engineering},
  volume  = {26},
  number  = {4},
  pages   = {59},
  year    = {2021},
  doi     = {10.1007/s10664-021-09959-3}
}

@article{balliu2023challenges,
  author={Balliu, Musard and Baudry, Benoit and Bobadilla, Sofia and Ekstedt, Mathias and Monperrus, Martin and Ron, Javier and Sharma, Aman and Skoglund, Gabriel and Soto-Valero, César and Wittlinger, Martin},
  journal={IEEE Security \& Privacy}, 
  title={Challenges of Producing Software Bill of Materials for Java}, 
  year={2023},
  volume={21},
  number={6},
  pages={12-23},
  doi={10.1109/MSEC.2023.3302956}
}

@inproceedings{imtiaz2021comparative,
  author    = {Imtiaz, Nasif and Thorn, Seaver and Williams, Laurie},
  title     = {A Comparative Study of Vulnerability Reporting by Software
               Composition Analysis Tools},
  booktitle = {Proceedings of the 15th ACM/IEEE International Symposium on
               Empirical Software Engineering and Measurement (ESEM)},
  articleno = {5},
  numpages  = {11},
  year      = {2021},
  publisher = {ACM},
  address   = {Bari, Italy},
  isbn      = {9781450386654},
  doi       = {10.1145/3475716.3475769}
}

@online{cisa_kev,
  author       = {{Cybersecurity and Infrastructure Security Agency}},
  title        = {Known Exploited Vulnerabilities Catalog},
  year         = {2026},
  url          = {https://www.cisa.gov/known-exploited-vulnerabilities-catalog},
}

@online{osv_database,
  author       = {{Google} and {OpenSSF}},
  title        = {{OSV.dev}: Open Source Vulnerabilities Database and {API}},
  year         = {2021},
  url          = {https://osv.dev/},
}

@online{githubsbom,
  author       = {{GitHub, Inc.}},
  title        = {{REST API} Endpoints for Software Bill of Materials ({SBOM})},
  year         = {2026},
  url          = {https://docs.github.com/en/rest/dependency-graph/sboms},
}

@online{awesome_agriculture,
  author       = {Johnston, Bryce},
  title        = {awesome-agriculture: Open Source Technology for Agriculture,
                  Farming, and Gardening},
  year         = {2018},
  howpublished = {GitHub repository},
  url          = {https://github.com/brycejohnston/awesome-agriculture},
}

@online{opensourceagriculture,
  author       = {Coleman, Guy},
  title        = {{OpenSourceAgriculture}: Collating Open-Source Datasets,
                  Software Tools and Deployment Platforms Related to
                  Open-Source Agriculture},
  year         = {2023},
  howpublished = {GitHub repository},
  url          = {https://github.com/geezacoleman/OpenSourceAgriculture},
}

@online{awesome_open_ag,
  author       = {{Common Garden}},
  title        = {awesome-open-ag: A List of Open Agriculture Related Projects
                  and Libraries},
  year         = {2017},
  howpublished = {GitHub repository},
  url          = {https://github.com/CommonGarden/awesome-open-ag},
}

@online{awesome_geospatial_companies,
  author       = {Rieke, Christoph},
  title        = {awesome-geospatial-companies: List and Map of Companies for
                  Geospatial Jobs},
  year         = {2020},
  howpublished = {GitHub repository},
  url          = {https://github.com/chrieke/awesome-geospatial-companies},
}

@article{benjamini1995controlling,
  author  = {Benjamini, Yoav and Hochberg, Yosef},
  title   = {Controlling the False Discovery Rate: A Practical and Powerful
             Approach to Multiple Testing},
  journal = {Journal of the Royal Statistical Society: Series B
             (Methodological)},
  volume  = {57},
  number  = {1},
  pages   = {289--300},
  year    = {1995},
  doi     = {10.1111/j.2517-6161.1995.tb02031.x}
}

@article{dunn1964multiple,
  author  = {Dunn, Olive Jean},
  title   = {Multiple Comparisons Using Rank Sums},
  journal = {Technometrics},
  volume  = {6},
  number  = {3},
  pages   = {241--252},
  year    = {1964},
  doi     = {10.1080/00401706.1964.10490181}
}

@article{mann1947test,
  author  = {Mann, Henry B. and Whitney, Donald R.},
  title   = {On a Test of Whether One of Two Random Variables is Stochastically
             Larger than the Other},
  journal = {The Annals of Mathematical Statistics},
  volume  = {18},
  number  = {1},
  pages   = {50--60},
  year    = {1947},
  doi     = {10.1214/aoms/1177730491}
}

@article{kruskal1952use,
  author  = {Kruskal, William H. and Wallis, W. Allen},
  title   = {Use of Ranks in One-Criterion Variance Analysis},
  journal = {Journal of the American Statistical Association},
  volume  = {47},
  number  = {260},
  pages   = {583--621},
  year    = {1952},
  doi     = {10.1080/01621459.1952.10483441}
}

@article{wilcoxon1945,
  author  = {Wilcoxon, Frank},
  title   = {Individual Comparisons by Ranking Methods},
  journal = {Biometrics Bulletin},
  volume  = {1},
  number  = {6},
  pages   = {80--83},
  year    = {1945},
  doi     = {10.2307/3001968}
}

@article{mahalanobis1936generalised,
  author    = {Mahalanobis, Prasanta Chandra},
  title     = {On the Generalised Distance in Statistics},
  journal   = {Proceedings of the National Institute of Sciences of India},
  volume    = {2},
  number    = {1},
  pages     = {49--55},
  year      = {1936}
}

@article{rubin1980bias,
  author  = {Rubin, Donald B.},
  title   = {Bias Reduction Using {Mahalanobis}-Metric Matching},
  journal = {Biometrics},
  volume  = {36},
  number  = {2},
  pages   = {293--298},
  year    = {1980},
  doi     = {10.2307/2529981}
}

@article{iacus2012causal,
  author  = {Iacus, Stefano M. and King, Gary and Porro, Giuseppe},
  title   = {Causal Inference without Balance Checking: Coarsened Exact
             Matching},
  journal = {Political Analysis},
  volume  = {20},
  number  = {1},
  pages   = {1--24},
  year    = {2012},
  doi     = {10.1093/pan/mpr013}
}

@article{austin2011introduction,
    author = {Peter C. Austin},
    title = {An Introduction to Propensity Score Methods for Reducing the Effects of Confounding in Observational Studies},
    journal = {Multivariate Behavioral Research},
    volume = {46},
    number = {3},
    pages = {399--424},
    year = {2011},
    publisher = {Routledge},
    doi = {10.1080/00273171.2011.568786},
    note ={PMID: 21818162},
    URL = { https://doi.org/10.1080/00273171.2011.568786},
    eprint = {https://doi.org/10.1080/00273171.2011.568786}
}

@article{stuart2010matching,
  author  = {Stuart, Elizabeth A.},
  title   = {Matching Methods for Causal Inference: A Review and a Look
             Forward},
  journal = {Statistical Science},
  volume  = {25},
  number  = {1},
  pages   = {1--21},
  year    = {2010},
  doi     = {10.1214/09-STS313}
}

@article{mackinnon1985some,
  author  = {MacKinnon, James G. and White, Halbert},
  title   = {Some Heteroskedasticity-Consistent Covariance Matrix Estimators
             with Improved Finite Sample Properties},
  journal = {Journal of Econometrics},
  volume  = {29},
  number  = {3},
  pages   = {305--325},
  year    = {1985},
  doi     = {10.1016/0304-4076(85)90158-7}
}


\end{document}